\documentclass[review]{elsarticle}

\usepackage{hyperref}
\usepackage{amsmath, amssymb, bm, mathtools} 
\usepackage{tikz, standalone, pgfplots} 
\usepackage{cleveref}
\usepackage{ulem}
\usepackage{subcaption}

\journal{IFAC Annual Reviews in Control}

\newdefinition{ex}{Example} 
\crefname{ex}{Example}{Examples}
\newdefinition{assum}{Assumption}
\crefname{assum}{Assumption}{Assumptions}

\crefname{thm}{Theorem}{Theorems}

\crefname{appendix}{Appendix}{Appendices}

\newcommand{\setreal}{\mathbb{R}}
\newcommand{\transp}{\mathsf{T}}
\newcommand{\col}{\mathrm{col}}

\newcommand{\Id}{I_d}
\newcommand{\vd}{v_d}

\newcommand{\trunot}[1]{#1}
\newcommand{\ytru}{\trunot{v}}
\newcommand{\wtru}{\trunot{w}}
\newcommand{\thetatru}{\trunot{\theta}}

\newcommand{\vtru}{\trunot{v}}

\newcommand{\estnot}[1]{{\hat{#1}}}
\newcommand{\yest}{\estnot{v}}
\newcommand{\west}{\estnot{w}}
\newcommand{\thetaest}{\estnot{\theta}}

\newcommand{\maxcondest}{\estnot{\mu}}

\newcommand{\vref}{v_r}

\newcommand{\relu}{\underline{\rho}}
\newcommand{\satfn}{\overline{\rho}}

\newcommand{\Iint}{I_{\text{int}}}
\newcommand{\Iext}{I_{\text{ext}}}

\newcommand{\uin}{u}
\newcommand{\fint}{g}
\newcommand{\Isynp}{I_{\rm{syn}, p}}
\newcommand{\Iion}{I_{\rm{ion}}}
\newcommand{\maxcond}{\mu}
\newcommand{\nernst}{E}

\newcommand{\INa}{I_{\text{Na}}}
\newcommand{\IK}{I_{\text{K}}}
\newcommand{\IL}{I_{\text{L}}}

\newcommand{\Na}{{\textrm{Na}}}
\newcommand{\K}{{\textrm{K}}}
\newcommand{\T}{{\textrm{T}}}
\newcommand{\Ca}{{\textrm{Ca}}}
\newcommand{\Hcond}{{\textrm{H}}}
\newcommand{\A}{{\textrm{A}}}
\newcommand{\Lcond}{{\textrm{L}}}
\newcommand{\KCa}{{\textrm{KCa}}}
\newcommand{\KIR}{{\textrm{KIR}}}
\newcommand{\GABA}{{\textrm{G}}}
\newcommand{\Leak}{{\textrm{leak}}}

\newcommand{\gain}{\gamma}

\newcommand{\taumin}{\underline{\tau}}
\newcommand{\taumax}{\overline{\tau}}

\newcommand{\ion}{{\rm{ion}}}
\newcommand{\syn}{{\rm{syn}}}
\newcommand{\gap}{{\rm{gap}}}
\newcommand{\pre}{{\textrm{p}}}

\begin{document}

\begin{frontmatter}

\title{Adaptive Conductance Control}


\author[uniaffiliation]{Raphael Schmetterling}
\author[uniaffiliation]{Thiago Burghi}
\author[uniaffiliation]{Rodolphe Sepulchre}


\address[uniaffiliation]{Department of Engineering, University of Cambridge, Trumpington Street, Cambridge CB2 1PZ, United Kingdom}

\begin{abstract}
Neuromodulation is central to the adaptation and robustness of animal nervous systems.
This paper explores the classical paradigm of indirect adaptive control to
design neuromodulatory controllers in  conductance-based neuronal models. 
The adaptive control of maximal conductance parameters is shown to provide
a methodology aligned with the central concepts of neuromodulation in physiology
and of impedance control in robotics.
\end{abstract}


\end{frontmatter}

\section{Introduction}

Due to the very nature of nervous electrical signals, brain medicine and brain-inspired computing put 
an increasing demand on completely novel control design methodologies. 
Neurotechnology interfaces neural tissues with electronic devices via the exchange of electrical signals. 
While of the
same physical nature,  those signals are radically different in animals and in machines:  neural signals are
analogue and spiky, whereas the information processing of actuator and sensor signals is
digital and quantised. Current control systems in neurotechnology 
use classical linear filtering techniques in combination with
 conventional analogue-to-digital converters from robotics and electronics \cite{sorrell_brainmachine_2021}. 
They fail to acknowledge the excitable dynamics underlying the generation of spiky signals.
Future developments  will call for more compliance and more integration
between the biological and technological domains. Such requirements will impose control interfaces
that interconnect (or, in the behavioural language of Willems \cite{willems_behavioral_2007}, {\it share}) signals of the same
nature.

The significance of designing a controller that shares the input-output properties of
the controlled physical system has a long and rich history in robotics. Passivity-based control is 
rooted in the concept of \textit{port interconnection}, which wires the physical terminals of a passive plant and a passive controller both
modelled as mechanical ports \cite{ortega_putting_2001}. In robotics, impedance control is rooted in the design of a controller that shapes the mechanical impedance
of the closed-loop system to comply with the environment, itself modelled in terms of a mechanical impedance \cite{hogan_impedance_1985}.

The present paper adopts the same philosophy for the control of a neuronal system. We use
the well-established framework of conductance-based modelling both
for the controller and for the system to be controlled. We consider conductance-based
neural networks in which each neural node is a one-port circuit composed of one leaky capacitor (the 'passive membrane') in parallel
with a bank of ohmic current sources of variable conductance. The controller is itself
an additional set of ohmic current sources connected in parallel to those of the neuron. Each conductance is voltage-dependent, gating
the current flow in a specific temporal and amplitude window.

A key emphasis in the present paper is on the adaptive control of the maximal conductances
of a conductance-based model. Each maximal conductance modulates the relative importance of a specific current source. Similar to the parameters of a conventional controller, the maximal conductances shape the total conductance of the controlled neurons. We wish to demonstrate that the online adaptation of
 maximal conductances provides a versatile framework to control the behaviour of a neuronal system. Online adaptation of the maximal conductances is aligned with the concept of \textit{neuromodulation}, which is of key
importance in the biological control of neuronal systems \cite{marder_neuromodulation_2014,sepulchre_control_2019,drion_ion_2015}. All nervous systems are subject to neuromodulation, and each neuron is potentially the target of multiple neuromodulators \cite{marder2012neuromodulation}. Furthermore, each modulator can act on multiple ionic currents in the same neuron \cite{marder2002cellular}.

Our methodology exploits
 the classical framework of model reference adaptive control \cite{astrom_adaptive_2008}. This framework relies on key physical properties of electrical and mechanical
 circuits: a relative degree one between the two terminal variables of the port (current and voltage in the present paper), and contracting internal dynamics. We
 show that these properties can be exploited in conductance-based electrical circuits in the same
 way as they have been exploited in the   impedance-based mechanical circuits of robotics.
A core element of the proposed adaptive control design is the adaptive observer
recently proposed for real-time estimation of conductance-based models in \cite{burghi2021online}.

The paper is organised as follows. Section \ref{sec:modelling} introduces conductance-based models, including the specific parametrisation we will require. Section \ref{sec:observer} summarises the adaptive observer design detailed in \cite{burghi2021online}.  Section \ref{sec:MRACC} employs this observer to solve the basic problems of adaptive reference tracking and adaptive disturbance rejection, as well as showing the relevance of such problems in the control of a simple biophysical neural network.  Section \ref{sec:discussion} discusses the idealised assumptions of the present paper and possible routes to make the theoretical methodology amenable to practical solutions of control problems in electrophysiology or in neuromorphic engineering.

\section{Conductance-Based Modelling}
\label{sec:modelling}

Since the seminal work of Hodgkin and Huxley \cite{hodgkin_quantitative_1952},
biophysical models in neurophysiology have been founded on nonlinear electrical circuits known as \textit{conductance-based models}. 
A detailed introduction to such models can be found in 
\cite{izhikevich_dynamical_2007,keener_mathematical_2009}. In this section, we extend the system-theoretic conductance-based modelling framework found in \cite{burghi2021online}.

\subsection{Conductance-based model of a neuron}

\begin{figure}[t]
	\begin{center}
		\includegraphics[width=\textwidth]{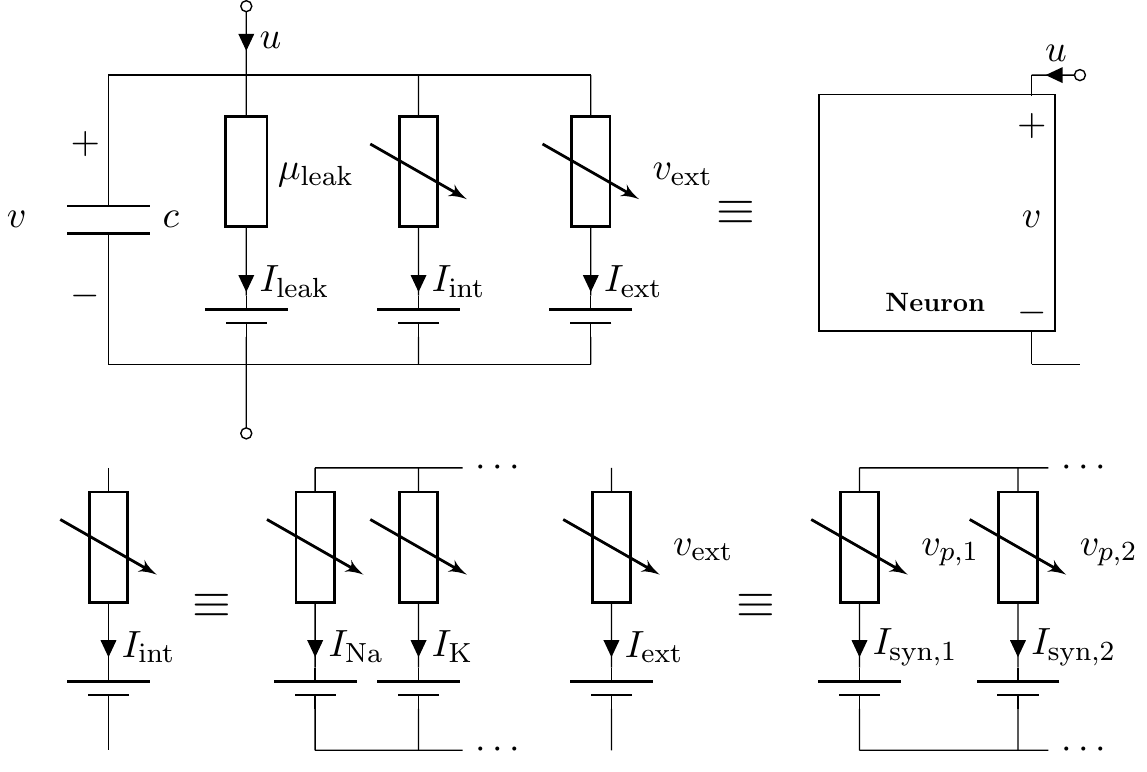}
	\end{center}
	\caption{Circuit representation of a neuron with voltage $v$ that is coupled
		with synapses 
		to presynaptic neurons $i$, each with voltage 
		$v_{p,i}$.}
	\label{fig:conductance_based}
\end{figure}

A conductance-based model of a neuron is a one-port electrical circuit 
with the architecture shown in Figure 
\ref{fig:conductance_based}: a capacitor of 
capacitance $c>0$ in parallel with 
a \textit{leak current} $I_{\Leak}$ and 
several \textit{intrinsic ionic currents}. The capacitance and leak current model the neuron's cell membrane, and the ionic currents model the flow of ions across this membrane.
Additional \textit{extrinsic} currents model the 
\textit{synaptic currents} $\Isynp$,
due to interconnections with other neurons, as
well as an \textit{input} (or \textit{applied})
\textit{current} $\uin$.
The latter represents the current injected into
the circuit by an intracellular
electrode.
The capacitor voltage $v$, which models the
neuronal membrane potential, evolves according to
Kirchhoff's law, that is
\begin{subequations}
\label{eq:general_cb}
\begin{align}
	\label{eq:single_neuron_cb}
	c \, \dot{v} &= 
	- I_{\Leak} 
	- \sum_{\ion \in \mathcal{I}} 
	\Iion
	-\sum_{\syn \in \mathcal{S}} 
	\sum_{\pre \in \mathcal{P}} 
	\Isynp
	+
	\uin \\
	&= - I_{\Leak} 
	- \Iint
	- \Iext
	+
	\uin
\end{align}
\end{subequations}
where $\Iint$ is the parallel combination of intrinsic ionic currents, $\Iext$ is the parallel combination of synaptic currents,
$\mathcal{I}$ is the index set of intrinsic ionic currents,
$\mathcal{S}$ is the index set of synaptic 
neurotransmitter types, and $\mathcal{P}$ is the
index set of \textit{presynaptic neurons}.

Each current in the circuit is ohmic in nature, 
but with a conductance that can be nonlinear and 
voltage-dependent. The leak current has a 
constant conductance and is given by
\[
I_{\Leak} = \maxcond_{\Leak} (v-\nernst_{\Leak}),
\]
whereas the intrinsic ionic currents are modelled by
\begin{subequations}
	\label{eq:current_cb}
	\begin{align}
		\label{eq:ion_currents}
		I_\ion &= \maxcond_\ion \, m_\ion^{p_\ion} \, h_\ion^{q_\ion} \, (v - \nernst_\ion) \\[.5em]
		\label{eq:activation}
		\tau_{m,\ion}(v) \dot{m}_\ion &= -m_\ion+ \sigma_{m,\ion}(v) \\[.5em]
		\label{eq:inactivation}
		\tau_{h,\ion}(v) \dot{h}_\ion &= -h_\ion+ \sigma_{h,\ion}(v)
	\end{align}
\end{subequations}
The constants $\maxcond_{\ion}>0$ 
and $\nernst_{\ion}\in\setreal$ are called
(intrinsic) \textit{maximal conductances} and 
\textit{reversal potentials}, respectively. 
The static \textit{activation functions} 
$\sigma_{m,\ion}(v)$
and 
$\sigma_{h,\ion}(v)$,
and
\textit{time-constant functions} $\tau_{m,\ion}(v)$ 
and $\tau_{h,\ion}(v)$, model the nonlinear gating 
of the ionic conductance. Because 
$\sigma_{m,\ion}:\setreal\to(0,1)$ 
and 
$\sigma_{h,\ion}:\setreal\to(0,1)$ 
are monotonically increasing and decreasing, respectively, the states $m_\ion$ and $h_\ion$ are called \textit{activation} and  \textit{inactivation gating variables}, respectively. The time-constant functions vary in shape, but always respect the bounds
\begin{equation*}
	0 < \taumin_\ion \le \tau_{m,\ion}(v),
	\tau_{h,\ion}(v) \le \taumax_\ion
\end{equation*}
for all $v \in \setreal$ and some $\taumin_\ion,\taumax_\ion > 0$. 
The exponents $p_\ion$ and $q_\ion$ in \eqref{eq:ion_currents} are
natural numbers (including zero).

\begin{ex}
	\label{ex:HH_description} 
	Conductance-based modelling originated in the
	seminal paper of	Hodgkin and Huxley  \cite{hodgkin_quantitative_1952}. 
    The Hodgkin-Huxley (HH) model
	includes two intrinsic ionic currents: a
	transient sodium current $\INa$ and
	a potassium current $\IK$, so that 
	$\mathcal{I} = \{\rm{Na},\rm{K}\}$. 
	The voltage dynamics of a single, isolated
	HH model (no synaptic currents) are given by 
	\[
		c \, \dot{v} = 	
		-\underbrace{\maxcond_{\rm{Na}} m_{\rm{Na}}^3 h_{\rm{Na}} 
			(v-\nernst_{\rm{Na}})}_{\INa} 
		-\underbrace{\maxcond_{\rm{K}} m_{\rm{K}}^4 
			(v-\nernst_{\rm{K}})}_{\IK} \\
		-\underbrace{\maxcond_{\Leak}(v-\nernst_\Leak)}_{I_{\Leak}}
		+ \uin
	\]	
	and the gating variable dynamics are of the form
	\eqref{eq:activation}-\eqref{eq:inactivation} for
	$\ion \in \{\Na,\K\}$.
\end{ex}

The HH model only includes two voltage-dependent conductances to parameterise the intrinsic current. Those types of currents are necessary and sufficient to model an action potential, or spike \cite{izhikevich_dynamical_2007}. This is due to the presence of fast-activating negative conductance and slower-activating positive conductance, which act as sources of positive and negative feedback respectively \cite{drion_neuronal_2015}. The fast negative conductance is provided by the activation of the inward current $\INa$ (due to the activation gating variable $m_{\rm{Na}}$) and the slower positive conductance is provided by  both the inactivation of $\INa$ (due to $h_{\rm{Na}}$) and the activation of the outward current $\IK$ (due to $m_{\rm{K}}$) \cite{sepulchre_excitable_2018,franci_balance_2013}.

Biophysical neurons may exhibit much richer behaviours, including for instance transitions between spiking and bursting patterns. The conductance-based models of such neurons include more conductances, leading to a plethora of single-neuron models differing from each other by the kinetics and activation ranges of their gating variables. 

\begin{ex}
	\label{ex:model_description} 

For the sake of illustration, all the neurons in the rest of this paper 
are modelled using the same conductance-based model, similar to the model in \cite{drion_switchable_2018}. 
This model includes eight distinct ohmic current sources, each modelling representative currents of
the experimental neurophysiological literature. Conductance control of such a model refers to adjusting the eight maximal parameter conductances 
of the model to shape the total intrinsic current of the neuron. We label those distinct currents with the index set  $\mathcal{I} = \{\Na,\Hcond, \T, \A, \K,\Lcond,\KCa,\KIR\}$. Some of these currents share the same type of ion, and so share the same reversal potential. The voltage dynamics of this model are as follows:
	\begin{equation*}
		\begin{split}
			c  \dot{v} = 
			&-\maxcond_{\Na} m_{\Na}^3 h_{\Na}(v-\nernst_{\Na})
			-\maxcond_{\Hcond} m_{\Hcond} (v-\nernst_{\Hcond}) \\
			&-\maxcond_{\T}m_{\T}^2 h_{\T}(v-\nernst_{\Ca})
			-\maxcond_{\A}m_{\A}^4 h_{\A} (v-\nernst_{\K}) \\
			&-\maxcond_{\K} m_{\K}^4 (v-\nernst_{\K})
			-\maxcond_{\Lcond} m_{\Lcond} (v-\nernst_{\Ca}) \\
			&-\maxcond_{\KCa} (\frac{[\Ca]}{15+[\Ca]})^4 (v-\nernst_{\K})
			-\maxcond_{\KIR} \sigma_{m,\KIR} (v-\nernst_{\K}) \\
			&
			-\maxcond_{\Leak}(v-\nernst_{\Leak}) + \uin
		\end{split}
	\end{equation*}
	Note the presence of the activation function $\sigma_{m,\KIR}$, implying that the gating  variable $m_\KIR$ is a static function of the voltage.
	Note also that the conductance of the current $I_\KCa$ depends on the calcium concentration $[Ca]$, rather than on the voltage. The calcium concentration is modelled as a low-pass
	filtered version of the voltage:
	\begin{equation}
		\label{eq:ca-dynamics}
		\dot{[Ca]} = -0.01 m_\Lcond (V - \nernst_{\Ca}) - 0.0025 [Ca]
	\end{equation}
	The remaining intrinsic gating variables evolve according to equations of the type  \eqref{eq:activation}-\eqref{eq:inactivation}. Figure \ref{fig:neuron_demo} illustrates the simulated behaviour of this model. The precise dynamics of each gating variable are 
	specified in \ref{sec:gating_dynamics}.
	
	\begin{figure}[t]
	\begin{center}
		\includegraphics[width=\textwidth]{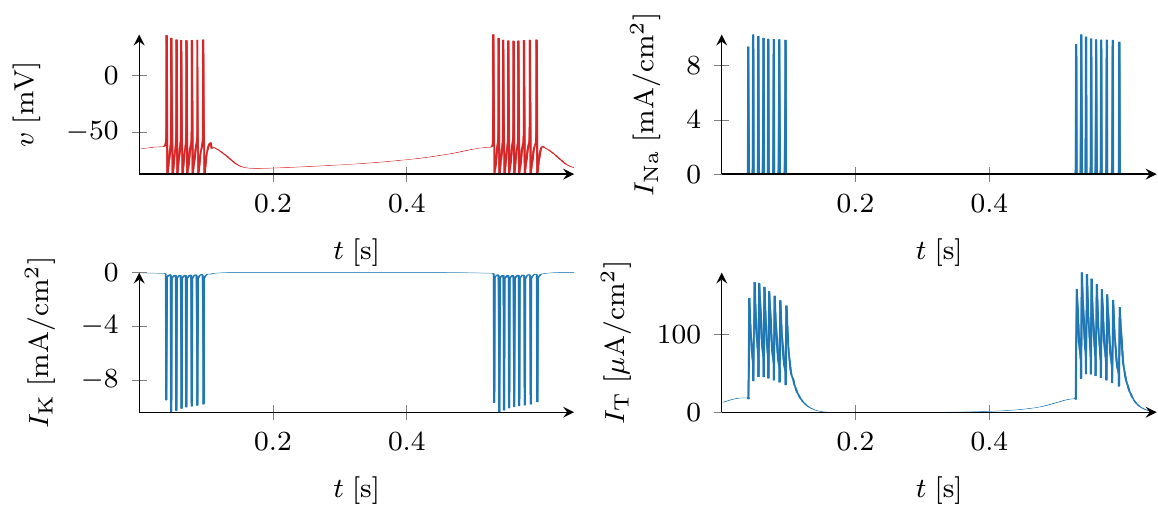}
	\end{center}
	\caption{Two bursts produced by the neuron model in \cref{ex:model_description}, showing the membrane potential $v$ (top left) and three selected ionic currents.}
	\label{fig:neuron_demo}
\end{figure}
\end{ex}

Recall that the neuron in Example \ref{ex:HH_description} exhibits spiking due to the presence of both fast-activating negative conductance and slower-activating positive conductance. The model in Example \ref{ex:model_description} exhibits robust bursting because this pairing is replicated at the timescale of the burst \cite{franci_robust_2017}. The T- and L- type calcium currents $I_\T$ and $\IL$ provide negative conductance in the slower timescale of calcium activation, and even slower positive conductance is provided by the calcium-activated potassium current $I_{\KCa}$.

Synaptic currents arise from electrochemical connections between neurons
\cite[Chapter 7]{ermentrout_mathematical_2010}. They are modelled as current sources in the same way as intrinsic currents;
the only difference is that the voltage dependence of their conductances is on the presynaptic neuron, noted $v_p$ in the example below.

\begin{ex}
	\label{ex:synapse_model_description} 
For the sake of illustration, all the models in this paper  only use one type of (inhibitory) synapse, called a GABA synapse in neurophysiology, obeying
the standard model reproduced from \cite{drion2018switchable}:
\begin{subequations}
	\label{eq:syn_dynamics}
	\begin{align}
		\label{eq:syn_current} 
		\Isynp &= \maxcond_{\syn,\pre} s_{\syn,\pre} \,
		(v - \nernst_\syn)
		\\[.5em]
		\label{eq:syn_activation} 
		\dot{s}_{\syn,\pre} &= 
		0.53\sigma_\syn(v_\pre)(1 - s_{\syn,\pre})
		-0.18s_{\syn,\pre}
	\end{align}
\end{subequations}
with a synaptic activation function $\sigma_\syn$ given by
a sigmoid function of the form
\begin{equation}
	\label{eq:synaptic_time-constant} 
	\sigma_\syn(v_\pre) = \frac{1}{
		1 + \exp(-(v_\pre - 2) / 5)}.
\end{equation}
Here, $s_{\syn,\pre}$ is the synaptic gating
variable and $v_\pre$ is the membrane voltage of the presynaptic neuron.
The constants $\maxcond_{\syn,\pre} > 0$ and 
$\nernst_\syn\in\setreal$ are (synaptic) maximal conductances and reversal
potentials, respectively. 
This model can represent excitatory or
inhibitory synapses, depending on the value of $\nernst_\syn$. In the present paper we set $\nernst_\syn=-90$,
which models the inhibitory synapses encountered in central pattern generators.

\end{ex}

\subsection{Conductance-based model of a neural network}
\label{sec:cb_networks}

A conductance-based neural network is given 
by the interconnection of $n_v \in \mathbb{N}$ single neurons. 
We denote the index set of neurons in the network by $\mathcal{N} := \{1,\dotsc,n_v\}$
and we describe the dynamics of the $i^{\rm{th}}$ 
neuron in the network by attaching an 
$i \in \mathcal{N}$ subscript to all the
variables in \eqref{eq:general_cb}-\eqref{eq:syn_dynamics}, except for reversal
potentials, time-constant functions, and 
activation functions. 

The network state-space model gathers all neuronal membrane 
voltages in the vector
\[
v = (v_1,\dotsc,v_{n_v})^\transp,
\]
and all other state variables in the vector
\[
w = \col(w^{(1)},\dotsc,w^{(n_v)}),
\]
where the vector $w^{(i)}$ collects the intrinsic and synaptic gating
variables of the $i^{\rm{th}}$ neuron, that is, $m_{\ion,i}$, $h_{\ion,i}$, 
and $s_{\syn,p,i}$, as well as the calcium concentration $[Ca]_i$.
Notice that with this notation, 
$\mathcal{P}\subseteq \mathcal{N}$.

In addition to synaptic current 
interconnections, neuronal networks may include
electrical gap junctions, modelled as 
ohmic resistive currents flowing between
neurons. Gap junction currents are thus 
passive components of the network, just as
leak currents are passive components in a
single neuron. It follows that the model of a
conductance-based neural network has a form
completely analogous to that of a single
neuron, given by \eqref{eq:general_cb}. 
The network model is given by the
vector equations
\begin{subequations}
\label{eq:general_cb_network}
\begin{align}
	\label{eq:single_neuron_cb_in_network}
	c \, \dot{v} 
	&= - I_{\Leak} 
	- \Iint
	- \Iext
	+
	\uin \\
	\label{eq:gating_vars_network}
	\dot{w} &= g(v, w)
\end{align}
\end{subequations}
where $I_{\Leak}$ is the overall leaky 
current, given by
\[
	I_{\Leak} = 
	\begin{pmatrix}
		\maxcond_{\Leak,1}(v_1
		-\nernst_{\Leak}) + 
		\sum_{i\in\mathcal{N}\backslash\{1\}} \maxcond_{\gap,1,i}
		(v_{1}-v_i)
		\\
		\vdots \\
		\maxcond_{\Leak,n_v}(v_{n_v}-
		\nernst_{\Leak}) +
		\sum_{i\in\mathcal{N}\backslash\{n_v\}} \maxcond_{\gap,n_v,i}
		(v_{n_v}-v_i)
	\end{pmatrix}
\]
and $\Iint \in \setreal^{n_v}$, 
$\Iext \in \setreal^{n_v}$ and 
$\uin\in \setreal^{n_v}$ are formed 
by gathering the intrinsic and extrinsic
currents and inputs of each neuron in the
corresponding $n_v$-dimensional vectors. The addition of gap junction currents extends the system-theoretic modelling framework of \cite{burghi2021online}.

The corresponding vector function $\fint(v,w)$ in \eqref{eq:gating_vars_network}, which represents 
the \textit{internal dynamics} of the neural network, collects the dynamics of all gating variables and calcium concentrations in the network, respecting the 
order in which those variables were included in $w$. For each 
neuron in the network, these dynamics are easily found from \eqref{eq:activation},
\eqref{eq:inactivation}, \eqref{eq:ca-dynamics} and \eqref{eq:syn_activation}.

\begin{ex}
	\label{ex:HCO_description} 
	A Half-Centre Oscillator (HCO) is a network of two neurons mutually 
	coupled by inhibitory synapses. 
	This elementary network is the simplest example of a \textit{central pattern generator},
	a type of neural network capable of generating autonomous rhythms for motor control \cite{marder_central_2001}.

The HCO model of this paper  interconnects two identical
	bursting neurons modelled according to
	\cref{ex:model_description} with
	the synapse model given in 
	\cref{ex:synapse_model_description}. The synapse index set is given by $\mathcal{S} = \{G\}$. 
	The state of the internal dynamics is given by $w = col(w^{(1)}, w^{(2)})$, with
	\begin{equation*}
		w^{(i)} = col(m_{\Na,i},h_{\Na,i},m_{\Hcond,i},m_{\T,i},h_{\T,i},m_{\A,i},
		m_{\K,i},m_{\Lcond,i},s_{j},[Ca]_i)
	\end{equation*}

	The HCO model is a building block of more complex central pattern generators such as the example considered in Section \ref{sec:network}, which includes gap junctions as well as synapses.
\end{ex}

\section{Adaptive observers for conductance-based models}
\label{sec:observer}

Following the notation of \cite{burghi2021online}, the conductance-based network model \eqref{eq:general_cb_network} 
can be written in the form
\begin{subequations}
	\label{eq:cb_true_system} 
	\begin{align}
		\label{eq:dy_true} 
		\dot{\vtru} &= 
		\Phi(\vtru,\wtru,\uin) \theta 
		+ b(\vtru,\wtru,\uin) \\
		\label{eq:dw_true} 
		\dot{\wtru} &= \fint(\vtru,\wtru)  \\
		\label{eq:dtheta_true}
		\dot{\thetatru} &= 0
	\end{align}
\end{subequations}
where $\vtru(t) \in \setreal^{n_v}$ is a vector of
measured output membrane voltages, 
$\uin(t) \in \setreal^{n_v}$ is a vector of
input currents, $\wtru(t)\in \setreal^{n_w}$ is a vector
of unmeasured internal states, and 
$\thetatru \in \setreal^{n_\theta}$ is a vector of
biophysical parameters to be estimated. The vector of unknown constant parameters $\thetatru$ is included in the state-space model via 
\eqref{eq:dtheta_true}.

In the present paper, the vector $\theta$ only includes maximal conductances.

\begin{ex}
	\label{ex:HCO_maxcond_parametrization} 
	Consider the HCO described in \cref{ex:HCO_description}. Let 
	\begin{equation}
		\label{eq:maxcond_parametrization_1} 
		\maxcond^{(i)} = (\maxcond_{\Na,i},\maxcond_{\Hcond,i},\maxcond_{\T,i}, 
		\maxcond_{\A,i}, \maxcond_{\K,i},\maxcond_{\Lcond,i},\maxcond_{\KCa,i},
		\maxcond_{\KIR,i}, \maxcond_{\syn,p,i},\maxcond_{\Leak,i})^\transp
	\end{equation}
	for $i,p\in \mathcal{N}=\{1,2\}$ and $p\neq i$. The estimated parameters of
	the HCO are chosen as
	\begin{equation}
		\label{eq:maxcond_parametrization_2} 
		\theta = \col(\maxcond^{(1)},\maxcond^{(2)})
	\end{equation}
	with $\maxcond^{(1)}$ and $\maxcond^{(2)}$ given by 
	\eqref{eq:maxcond_parametrization_1}.
	Letting $v=(v_1,v_2)^\transp$ and $w = \col(w^{(1)},w^{(2)})$, the voltage dynamics 
	of the model can then be written as \eqref{eq:dy_true}, where
	\[		
	\begin{split}
		\Phi(\vtru,\wtru) &=
		\begin{bmatrix}
			\varphi(v_1,w^{(1)}) & 0 \\
			0 & \varphi(v_2,w^{(2)})
		\end{bmatrix} \\
		b(t) &= (\uin_1(t)/c_1, \uin_2(t)/c_2)^\transp
	\end{split}
	\]
	with
	\[
	\varphi(v_i,w^{(i)}) = 
	-\frac{1}{c_i}
	\begin{pmatrix}
		m_{\Na,i}^3 h_{\Na,i}(v_i-\nernst_{\Na})
		\\[.5em]
		m_{\Hcond,i} (v_i-\nernst_{\Hcond})
		\\[.5em]
		m_{\T,i}^2 h_{\T,i}(v_i-\nernst_{\Ca})
		\\[.5em]
		m_{\A,i}^4 h_{\A,i} (v_i-\nernst_{\K})
		\\[.5em]
		m_{\K,i}^4 (v_i-\nernst_{\K})
		\\[.5em]
		m_{\Lcond,i} (v_i-\nernst_{\Ca})
		\\[.5em]
		(\frac{[\Ca]_i}{15+[\Ca]_i})^4 (v_i-\nernst_{\K})
		\\[.5em]
		\sigma_{m,\KIR}(v_i) (v_i-\nernst_{\K})
		\\[.5em]
		s_{\syn,\pre,i}(v_i-\nernst_\GABA)
		\\[.5em]
		v_i-\nernst_\Leak
	\end{pmatrix}^\transp
	\]
	for $i=1,2$ and $p\neq i$.
\end{ex}

An important property of the parametrisation in
\cref{ex:HCO_maxcond_parametrization} is that it
is decentralised: the network estimation problem
decouples into independent single-neuron
estimation problems. This decoupling allows the
estimation problem to be scaled to a possibly high-dimensional
network.

The recent work \cite{burghi2021online} provides an adaptive observer to estimate the parameters of the system (\ref{eq:cb_true_system}) in real-time.
This observer has global convergence properties and is based on the recursive least squares algorithm. The state-space
realisation of the observer is 
\begin{subequations}
	\label{eq:adaptive_observer}
		\begin{align}
					\label{eq:dy} 
			\dot{\yest} &= 
			\Phi(\ytru,\west,\uin)\thetaest 
			+ b(\ytru,\west,\uin)
			+ \gain(I +\Psi P \Psi^\transp) 
			(\ytru-\yest)
			\\
				  	\label{eq:dw}
			\dot{\west} &= \fint(\ytru,\west) 
			\\
				  	\label{eq:dtheta}
			\dot{\thetaest} &= \gain P \, 
			\Psi^\transp \, (\ytru-\yest)
			\end{align}
\end{subequations}
where $\gain>0$ is a constant gain, and the matrices $P$ and $\Psi$ 
evolve according to
\begin{subequations}
	\label{eq:adaptive_matrices} 
	\begin{align}
		\label{eq:dPsi}
		\dot{\Psi} &= 
		- \gain \Psi + 
		\Phi(\ytru,\west,\uin),
		\\
		\label{eq:dP} 
		\dot{P} &= \alpha P -  
		P \, \Psi^\transp \Psi P, & P(0) \succ 0
	\end{align}
\end{subequations}
where $\alpha>0$ is a constant forgetting rate, required to discount the initial error between $\wtru(0)$ and $\west(0)$. It can be shown that this adaptive observer recursively solves a least-squares regression problem with exponential forgetting, where the regression error is defined by filtering the derivative of $v$ with the first-order filter $H=\gamma/(s+\gamma)$. Without loss of generality, we assume $\Psi(0) = 0$.

The convergence of the above adaptive observer relies on the key property that the internal dynamics \eqref{eq:dw_true} are contracting in $w$ on a positively invariant compact set, uniformly in the voltage $v$. 
We refer the reader to  \cite{burghi2021online} for further details, including 
a contraction-based proof of convergence. 

\section{Model Reference Adaptive Conductance Control}
\label{sec:MRACC}
\begin{figure}
	\begin{center}
		\includegraphics[width=\textwidth]{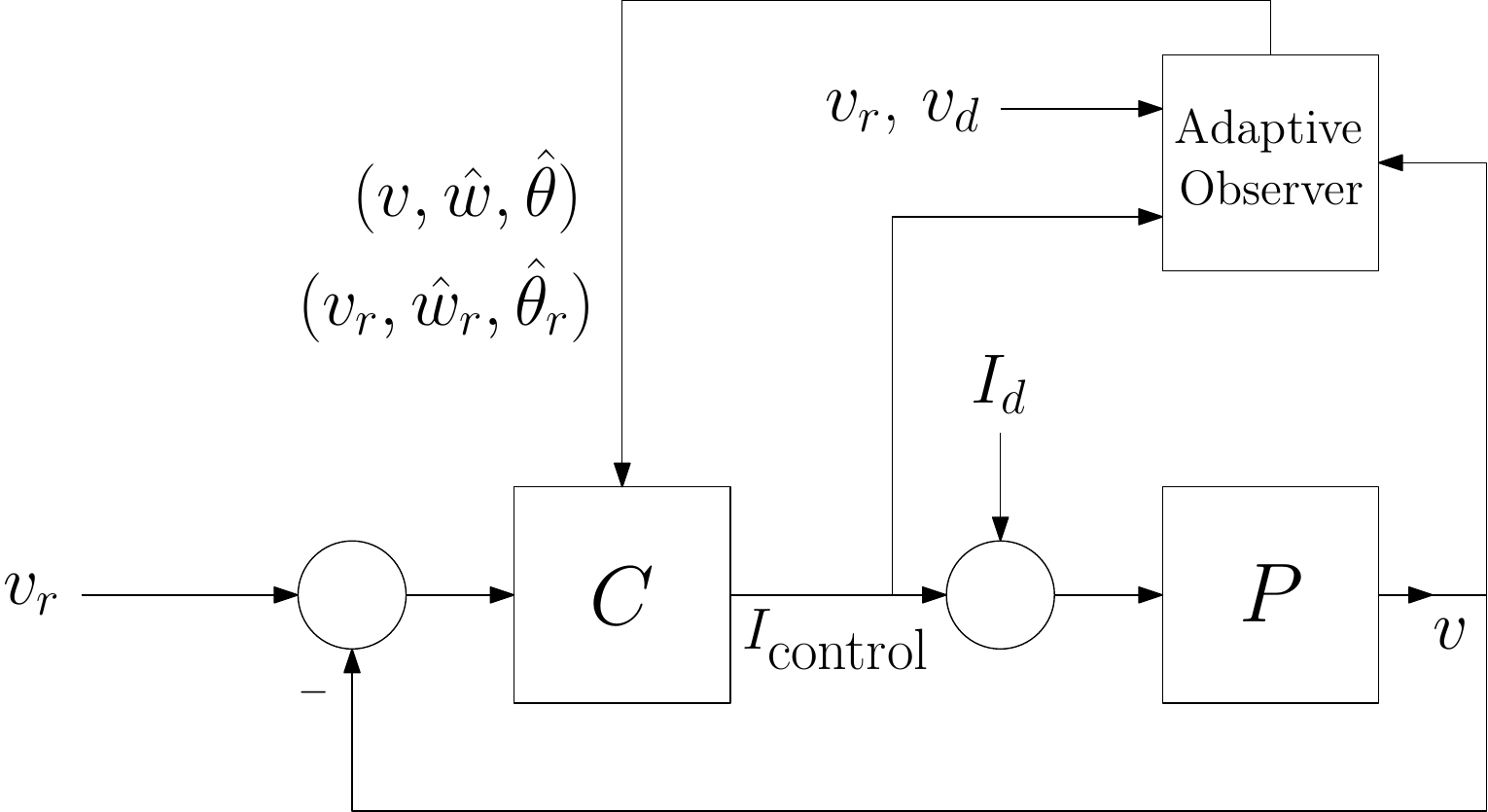}
	\end{center}
	\caption{Block diagram representation of the adaptive reference tracking and disturbance rejection problems. Adapted from \cite[Chapter 1]{astrom_adaptive_2008}.}
	\label{fig:block-diag}
\end{figure}

Adaptive observers are instrumental to the classical  design approach called {\it model reference adaptive control  (MRAC)} \cite{astrom_adaptive_2008}. In this section we illustrate
the application of model reference adaptive control to conductance-based models. We regard a single neuron as a voltage-controlled circuit.  We review the two canonical control problems of adaptive control: the adaptive tracking of a reference signal $\vref$ (Section \ref{sec:tracking}), and the adaptive disturbance rejection of an external current $I_{d}$ (Section \ref{sec:rejection}). We then illustrate the relevance of those elementary control problems in a network example (Section \ref{sec:network}). See Figure \ref{fig:block-diag} for a block diagram representation of the two problems.

\subsection{Adaptive reference tracking}
\label{sec:tracking}

The classical problem of reference tracking is to design a controller such that  the voltage output $v$ asymptotically converges to
the voltage reference $\vref$. If we assume that both the output $v$ and the reference $\vref$ are solutions of identical conductance-based models, this tracking problem reduces to the classical problem of {\it synchronisation}. 

Here we consider the  {\it adaptive} version of the tracking problem: we assume that the reference generator is 
a conductance-based model with constant but unknown vector parameter $\theta_r$ and that the controlled circuit is
a conductance-based model with the same model structure, and with a parameter $\theta$ that is also unknown. The problem of adaptive synchronisation has been
considered in \cite{fradkov_adaptive_1997,fradkov_adaptive_2000}, in the context of secure communication via encryption
in chaotic reference generators. Assuming a perfect knowledge of the ion channel kinetics $g(v,w)$, the solution presented here is relevant to track the neuromodulation
of a neuron in vivo or to learn the parameters of a silicon neuron from experimental traces.

We will first describe how to solve adaptive reference tracking without synchronisation. This yields an oscillation which follows the reference but with a possible phase shift. The phase shift is then eliminated with an additional resistive element in the controller.

We solve the adaptive reference tracking problem by estimating both the reference model and the controlled model 
with the observer design of the previous section. The observer of the reference generator provides an estimate $\hat{x}_r$
of the reference state vector $x_r = (\vref, w_r, \theta_r)$
whereas the observer of the controlled model provides an estimate  $\hat x$ of the controlled model state vector $x=(v,w,\theta)$. We use the control law $u(t) = I_\textrm{control}(t) + u_r(t)$, where $u_r(t)$ is the input to the reference neuron, and
\begin{equation}
	\label{eq:control_law_nondiffusive} 
	I_\textrm{control}(t) = 
	\Phi(v,\hat{w}) (\relu(\hat{\theta}_r) - \satfn(\hat{\theta})),
\end{equation}
where $\relu(x) \coloneqq \max{(0,x)}$ is the rectified linear (ReLU) function, and $\satfn(x) \coloneqq \min(x,\beta)$. Both functions are applied to their arguments elementwise. Together, they ensure that the solutions of the controlled neuron remain in a positively invariant set, as required for convergence of the observer \cite{burghi2021online}. We require $\beta$ to satisfy $\max_j\{\theta_j\} \leq \beta \leq \bar{\beta}$, where $\bar{\beta}$ is the largest value which preserves the set. We empirically choose $\beta$ by setting it to a large value (relative to plausible values of $\theta$) and reducing it if the membrane potential of the controlled neuron diverges.

Exponential convergence of the estimated parameters ($\hat{\theta}$ and $\hat{\theta}_r$) to the true parameters provides a solution to the tracking problem without synchronisation. The proof relies on the \textit{virtual system} idea of contraction theory \cite{lohmiller_contraction_1998}, and follows the same lines as in 
\cite[Section V.C]{burghi2021online}: the estimate $\hat{x}_r$ contracts exponentially fast to the reference $x_r$ while the estimate $\hat x$
contracts exponentially fast to $x$. Upon convergence of both the reference and plant observers, we obtain the non-adaptive synchronisation problem between two identical systems.

The solution to this problem
is particularly simple for conductan\-ce-based models because of their property of output feedback contraction or output
feedback incremental passivity \cite{wang_partial_2004,stan_analysis_2007}. Contraction of the error dynamics is ensured by including the output feedback term $\kappa (v_r - v)$ in the control law, for a sufficiently large gain $\kappa >0$. The circuit realisation of this feedback term is a resistive
wire between the reference and controlled circuit, or a {\it gap} junction in the language of neurophysiology. When this wire is introduced, Theorem 2 of \cite{wang_partial_2004} applies, and $v(t) \to \vref(t)$ as $t \to \infty$. The full control law is thus given by
\begin{equation}
	\label{eq:control_law_diffusive} 
	u(t) =I_\textrm{control}(t) + \kappa (v_r(t) - v(t)) + u_r(t),
\end{equation}
where $I_\textrm{control}(t)$ is given by \eqref{eq:control_law_nondiffusive}.

\begin{figure}
	\begin{center}
			\includegraphics[width=\textwidth]{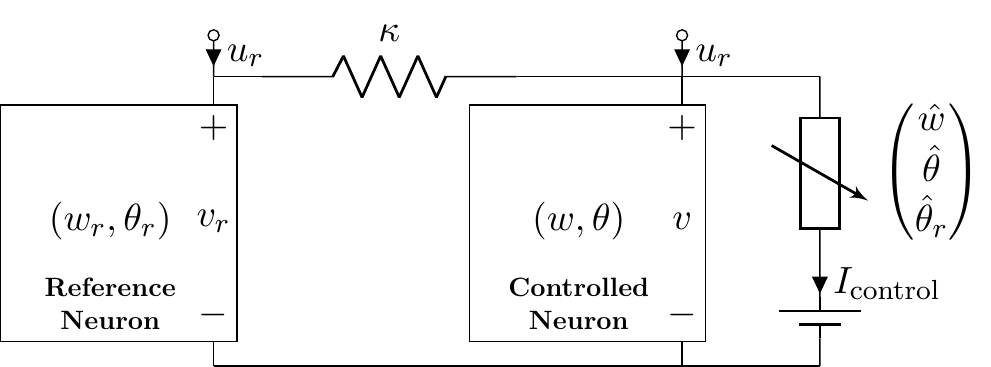}
		\end{center}
	\caption{Circuit diagram of the adaptive reference tracking problem.}
	\label{fig:reftrack-circ}
\end{figure}

\begin{figure}
    \begin{minipage}[b]{\linewidth}
	\begin{center}
		\includegraphics[width=\textwidth]{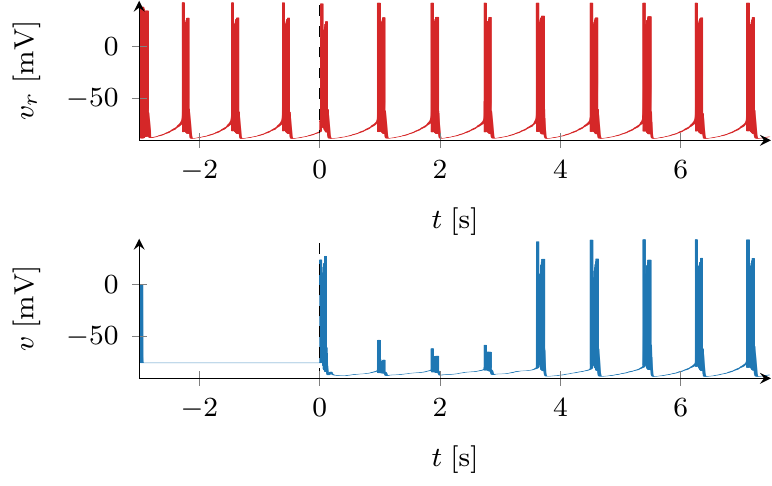}
	\end{center}
	\subcaption{Top (resp. bottom): $\vref$ (resp. $v$) before and during the running of the observers and controller. The observers, controller and coupling are introduced at $t = 0$.}
	\end{minipage}
	\begin{minipage}[b]{\linewidth}
	\begin{center}
		\includegraphics[width=\textwidth]{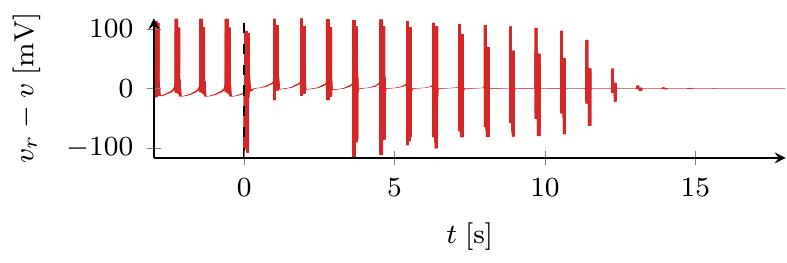}
	\end{center}
	\subcaption{Reference tracking error, including for $t < 0$ when there is no observer, controller or coupling.}	
	\end{minipage}
	\caption{Simulation of single-neuron reference tracking.}
	\label{fig:reftrack_plots}
\end{figure}

Figure \ref{fig:reftrack_plots} illustrates the performance of the adaptive conductance control in a scenario where both the reference and controlled neurons
are the bursting neurons of Example \ref{ex:model_description}. To simplify the exposition, the parameters of the controlled neuron are set at $\theta=0$, meaning that the 'open-loop' controlled
neuron is the model of a passive membrane. The control scheme is shown in Figure \ref{fig:reftrack-circ}. The observers and the controller are switched on at time $t = 0$, after which the parameters of the controlled
neuron converge to the parameter values of the reference neuron. See \ref{sec:model_parameters} for the parameters used.

\subsection{Adaptive disturbance rejection}
\label{sec:rejection}

The classical problem of disturbance rejection has a solution similar to that of the tracking problem. 
Given a disturbance current $I_{d}$ generated by a synaptic current, we wish to design a feedback controller 
that asymptotically rejects that disturbance.   A classical solution of the disturbance rejection problem is to include a model of the disturbance 
in the controller. 

This problem is of relevance in electrophysiology. Electrophysiologists study the properties of
a given neuronal circuit \textit{in vitro} by
extracting the circuit from its nervous system
and probing its responses to electrical stimuli
in an experimental preparation, see e.g. \cite{prinz_dynamic_2004}.
Classical solutions include  pharmacological agents that block specific types of 
 ion channels, thereby reducing the synaptic currents to zero;
see e.g. \cite{golowasch_ionic_1992}. A 
downside of using pharmacological agents is that
they may affect other properties of the circuit; 
another downside is the global
effect of such agents within a given preparation. 
These may be undesirable when
highly specific ion channel blocking is 
required.

An interesting alternative for targeted synaptic
isolation in an experimental preparation is the
design of a conductance-based controller for
the rejection of synaptic currents. In this case,
synaptic currents to be blocked are regarded as 
disturbances to be rejected.

\begin{figure}
	\begin{center}
		\includegraphics[width=\textwidth]{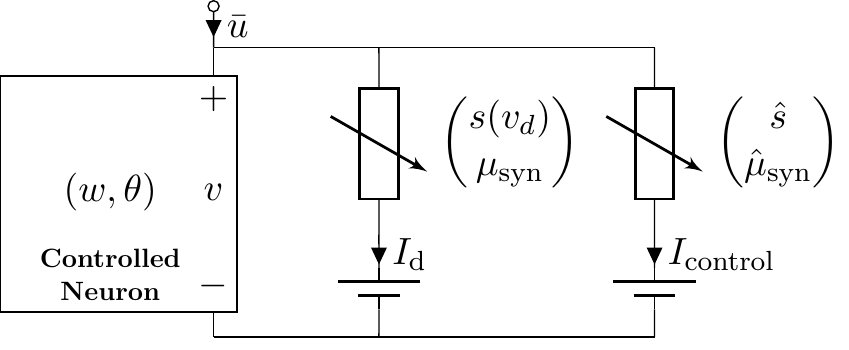}
	\end{center}
	\caption{Circuit diagram of the disturbance rejection problem, including the required controller.}
	\label{fig:distrej-circ}
\end{figure}

Assuming that the pre- and post-synaptic  voltages $v_d$ and $v$ 
are measured, a target 
synaptic current flowing between the two neuronal membranes
can be blocked by means of the adaptive disturbance 
rejection control scheme shown in Figure \ref{fig:distrej-circ}.
Note that this is the same circuit as in Figure \ref{fig:conductance_based}, but with two additional circuit elements connected in parallel. 

The first additional element is the disturbance $\Id$, which is interpreted as the specific synaptic current to be blocked. 
This disturbance current is modelled as
\begin{subequations}
	\label{eq:Id}
	\begin{align}
		&\Id = -\maxcond_{\textrm{syn}} s(\vd) (v - E_\textrm{syn}) \\
		&\dot{s} = 
		a_1 \sigma_s(v_d) (1 - s)
		- a_2 s, \label{eq:Ids}
	\end{align}
\end{subequations}
where $\sigma_s$ is a monotonically increasing activation function and $a_1, a_2 > 0$ are constant (known) synaptic parameters.

The second additional element is the controller.
Its inputs are the measured voltages $v$ and $\vd$, and
it generates a control current $I_\textrm{control}$ which
is designed to cancel $\Id$. We require that the behaviour of the closed-loop circuit converges to that of
an undisturbed conductance-based model, that is, one where the
targeted synaptic connection is absent.
We assume that both the pre- and post-synaptic neurons are the bursting neurons of
Example \ref{ex:model_description}, and that they are interconnected
with the inhibitory synapse of Example \ref{ex:synapse_model_description}.

To achieve perfect disturbance rejection, the unknown synaptic
maximal conductance $\maxcond_{\textrm{syn}}$ has to be estimated;
this estimate is denoted by $\maxcondest_{\textrm{syn}}$. 
The disturbance rejection controller can then be designed following
the \textit{certainty equivalence principle}
\cite{astrom_adaptive_2008}. In other words, using the estimate
$\maxcondest_{\textrm{syn}}$, the controller is designed to perfectly
cancel $\Id$ when 
$\maxcondest_{\textrm{syn}}=\maxcond_{\textrm{syn}}$. In this case,
the estimation problem therefore requires an estimation method
such that $\maxcondest_{\textrm{syn}} \to \maxcond_{\textrm{syn}}
\;\; \text{as} \;\; t \to \infty$. This can be accomplished by the 
observer given by \eqref{eq:adaptive_observer}-\eqref{eq:adaptive_matrices}, 
using the parametrisation $\hat{\theta} = \maxcondest_{\textrm{syn}}$. The observer also produces an estimate of the synaptic gating variable such that $\hat{s} \to s \;\; \text{as} \;\; t \to \infty$.
Given this observer, the input to the neuron is given by $u = I_\textrm{control} + \Id + \bar{u}$, where $\bar{u}$ is an arbitrary new input signal, and
\begin{subequations}
	\label{eq:Icontrol_DR}
	\begin{align}
		I_\textrm{control} &= -\hat{I}_d \\
		\hat{I}_d &= -\satfn(\maxcondest_\textrm{syn}) \hat{s} (v - E_\textrm{syn}) \\
		\dot{\hat{s}} &= 
		a_1 \sigma_s(v_d) (1 - \hat{s})
		-a_2 \hat{s},
	\end{align}
\end{subequations}
where $\sigma_s$, $a_1$ and $a_2$ are the same as in \eqref{eq:Ids}. As in Section \ref{sec:tracking}, the function $\satfn$ ensures the positively invariant set.

\begin{figure}
	\begin{minipage}[b]{\linewidth}
		\begin{center}
			\includegraphics[width=\textwidth]{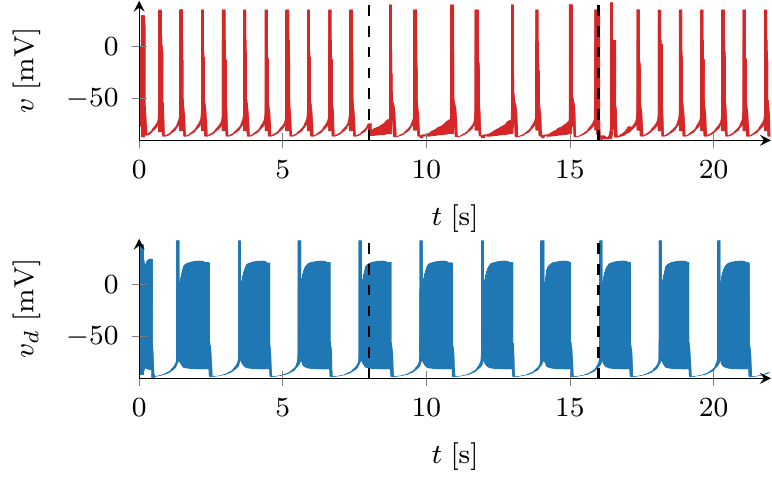}
		\end{center}
		\subcaption{Top (resp. bottom): $v$ (resp. $v_d$) during three stages of the experiment. For $t < 8$, the neuron is undisturbed (the synapse is not present). For $8 < t < 16$, the synapse is present but there is neither observer nor controller. The observer and controller are introduced at $t = 16$.}\label{fig:distrej-a}
	\end{minipage}
	\begin{minipage}[b]{\linewidth}
		\begin{center}
			\includegraphics[width=\textwidth]{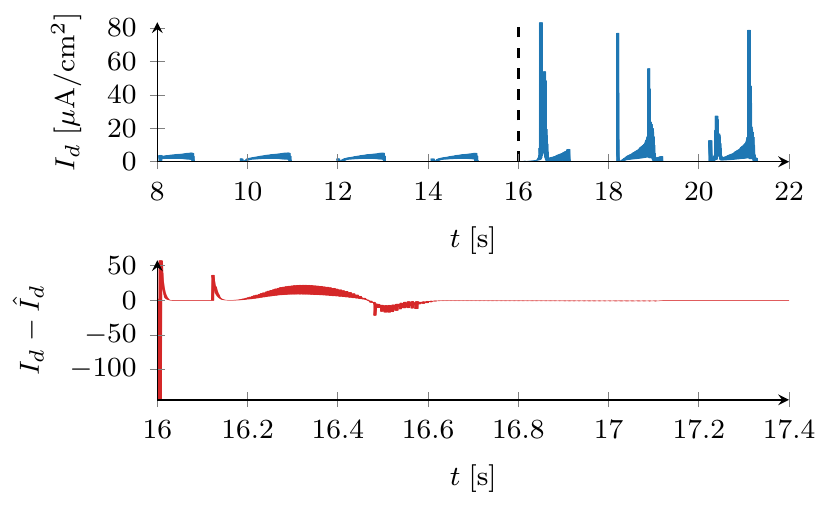}
		\end{center}
		\subcaption{Top: True value of the disturbance synaptic current, $I_d$. This is larger when $t > 16$ as the bursts of the two neurons have more overlap ($\Id$ is a function of both $v$ and $v_d$). Bottom: The observer's error in estimating $I_d$, defined by the difference between the true value and the estimate $\hat{I}_d$. The plot is truncated once the estimate has converged to the true value.}\label{fig:distrej-b}
	\end{minipage}
	\caption{Simulation of single-neuron disturbance rejection.}\label{fig:plots-distrej}
\end{figure}

Figure \ref{fig:plots-distrej} illustrates the performance of the
disturbance rejection controller. The model and observer parameters,
and the input currents, are provided in \ref{sec:model_parameters}.

\subsection{Network neuromodulation}
\label{sec:network}

The adaptive conductance controller developed in the previous sections for a single cell 
is by nature decentralised: it can be applied independently to different neurons (nodes) 
in a network. Each neuron in the network 
receives synaptic currents that can be adaptively estimated using the measurements of the presynaptic
neuron voltages. As a consequence, an observer can be designed for each neuron of the network,
and each of those neurons can be controlled to synchronise to a given neuron or to
adaptively reject specific synaptic currents. 

\begin{figure}
	\begin{minipage}[b]{\linewidth}
		\begin{center}
			\includegraphics[width=\textwidth]{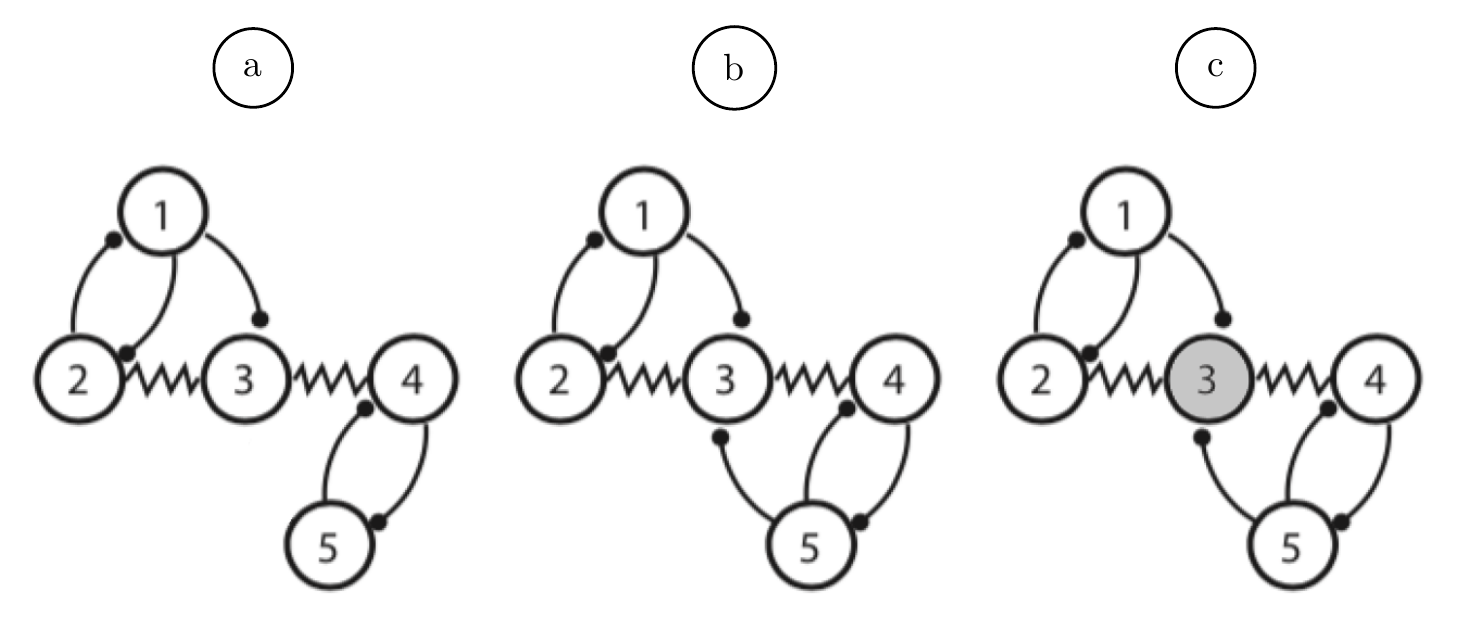}
		\end{center}
		\subcaption{The network simulation is divided into three phases: (a) The disturbance (defined as the synapse from Neuron 5 to Neuron 3) is not present. (b) The disturbance is introduced. (c) The observer and controller are introduced (Neuron 3 is coloured grey to indicate that it is the controlled neuron). Diagram adapted from \cite{drion_cellular_2019}.}\label{fig:orchestron_diags}
	\end{minipage}
	\begin{minipage}[b]{\linewidth}
		\begin{center}
			\includegraphics[width=\textwidth]{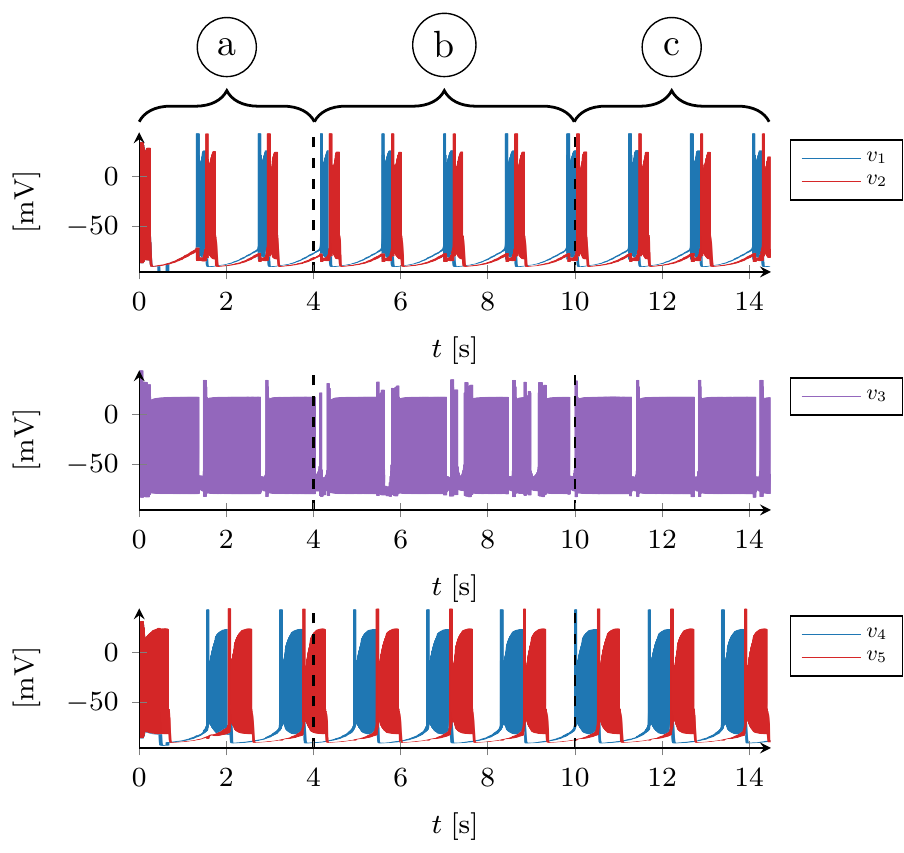}
		\end{center}
		\subcaption{Top: First HCO pair. Middle: Hub neuron. Bottom: Second HCO pair. The simulation is divided into the three phases defined in Figure \ref{fig:orchestron_diags}. In phase (a), $v_3$ is a simple bursting oscillation and is inhibited only when $v_1$ is active.
		In (b), which starts at $t = 4$, the hub expresses a 'mixed' rhythm as it is inhibited by both $v_1$ and $v_5$. This causes the bursts of $v_3$ to vary in length.
		During (c), starting at $t = 10$, the bursting behaviour converges to the same as in (a).}\label{fig:plots_orchestron}
	\end{minipage}
	\caption{Network disturbance rejection.}\label{fig:orchestron}
\end{figure}

We illustrate the versatility of this adaptive conductance control in a five-neuron network previously
analysed in \cite{drion_cellular_2019} and itself inspired by the Stomatogastric Ganglion, a 
crustacean central pattern generator \cite{marder2012neuromodulation}. 
The network interconnects a fast HCO and a slow HCO, both with the model structure of 
\cref{ex:HCO_description,ex:HCO_maxcond_parametrization}, 
through a central "hub" neuron. The connectivity
diagram of this network is shown in the centre of Figure \ref{fig:orchestron_diags}. 
Notice the lack of direct connections between the two HCOs on either side of the hub neuron.

In previous work \cite{drion_cellular_2019}, we have shown that this network can be switched between distinct rhythmic
states by the modulation of specific internal conductances. In every possible rhythmic configuration,
the network is composed of five neurons generated with the model of Example \ref{ex:model_description} and interconnected
with gap junctions and the inhibitory synapses of Example \ref{ex:synapse_model_description}. In an 
application of model-reference adaptive control, the challenge is to adaptively regulate the network
by only modulating the maximal conductance parameters. This can involve up to five distinct observers, assuming measurement of the five neuronal voltages.

As a simple illustration, we show how to decouple the two central pattern generators
by disconnecting the central hub using the disturbance rejection controller of the previous
section. This is achieved by control of the hub (Neuron 3 in Figure \ref{fig:orchestron_diags}). The controller rejects the inhibitory synapse from Neuron 5, using the same control law as Section \ref{sec:rejection}. This is illustrated in Figure \ref{fig:plots_orchestron} by the behaviour of the hub's membrane potential, $v_3$. During part (a) of the simulation,  $v_3$ expresses a rhythm governed almost entirely by the first HCO (that is, Neurons 1 and 2). As only Neuron 1 is inhibiting the hub, $v_3$ is low when $v_1$ is active and each burst of $v_3$ is the same length. In part (b), when the disturbance is introduced, $v_3$ expresses a 'mixed' rhythm governed equally by both HCOs. Bursts of $v_3$ are interrupted whenever $v_1$ or $v_5$ is active. Finally, the observer and controller are introduced in part (c), and $v_3$ converges to the undisturbed rhythm of (a). The simulation parameters are provided in \ref{sec:model_parameters}.

\section{Discussion}
\label{sec:discussion}

The results in Section \ref{sec:MRACC} suggest that Model Reference Adaptive Control, a classical control paradigm, provides a sound
methodological framework for designing online neuromodulation in conductance-based neuronal networks. The control strategy is
neuromorphic in the sense that it imitates the continuous adaptation of maximal conductances in biological networks by 
neuromodulators. It is also classical in the sense that it closely resembles the application of adaptive control in robotics. 

The simulations presented in Section \ref{sec:MRACC} are of course highly idealised: they assume a perfect modelling knowledge of the
voltage-dependent conductances and ideal measurements of all cellular voltages. These  assumptions are certainly not met in any
practical environment, whether in an experiment of electrophysiology or in the design of analogue neuromorphic hardware. Both in vivo
and in silico, the dynamics of gating variables are both uncertain and variable, that is, they vary with the environment (e.g. temperature) and they vary
from cell to cell. Likewise, voltage measurements are noisy, both in vivo and in silico, which introduces trade-offs and constraints 
in the observer parameters. For a rigorous study of how noise impacts the system identification of 
conductance-based models, see \cite{burghi2021feedback}. 
Robustness of the adaptive control laws against modelling uncertainty and measurement noise will
be the scope of future investigation, both analytical and experimental.

We envision at least three distinct angles of attack to increase the robustness and the biological plausibility of the proposed adaptive
design:
\begin{itemize}
\item 
{\bf Redundancy and degeneracy} of conductances seem essential in biological neurons to cope with uncertainty. Biological neurons
can use vastly different choices of maximal conductances to exhibit the same behaviour \cite{marder_neuromodulation_2014}. It has been suggested that this degeneracy plays 
a key role in homeostasis \cite{OLeary2014}. Viewing the internal dynamics of a neuron as a bank of nonlinear filters that collectively shape
the conductance of the total internal current, the adaptive controller functions very much like a one-layer artificial neural network with recursive
least-squares estimation of its parameters (this provides a link between biophysical models and the phenomenological models proposed in \cite{burghi_system_2020}). 
The bank of conductances is however not arbitrary in neurophysiology. Ion channel {\it types} identify specific
time scales and amplitude ranges of activation that are critical for the excitability thresholds of the neuron. These specific scales of activation 
have been well-documented over a long history of voltage-clamp experiments. The resulting dynamic conductances shape the closed-loop 
behaviour very much like the zeros and poles of a classical LTI controller shape the sensitivity of the closed-loop system \cite{drion_neuronal_2015,franci_realization_2014}.

\item
{\bf Voltage measurements} are assumed to be exact in this paper, but the spiking nature of the signals suggests that a much coarser information
about reference or disturbance signals might be sufficient to modulate the rhythm of a neuron. Future work will explore this possibility to reduce
the computational demand of the full observer. Here, neurophysiology will also be a useful guide. For instance, it is well-known that calcium is 
an essential second messenger involved in neuromodulation. In the present paper (as in many models of the literature), the intracellular 
calcium concentration is simply modelled as a low-pass filtered version of the voltage (as in \eqref{eq:ca-dynamics}). Earlier models have suggested simple yet
general homeostatic principles for the adaptation of cellular conductances using the intracellular calcium concentration as a feedback signal \cite{OLeary2014}.

\item
{\bf Hierarchical adaptation.} The adaptive controller in this paper is at the cellular scale, and the emphasis is on showing that adaptive controllers
at the cellular scale can modulate behaviours at the network level. But there is no doubt that adaptation is multi-scale in biological networks. 
Adaptive conductance control of conductance-based models is conducive to hierarchical and multi-scale controllers that deserve further
research. Adaptive control of a synaptic conductance can be further decentralised and use a conductance observer that only involves
the synaptic gating variable, as well as the pre- and post- synaptic voltages. Likewise, adaptive control of a mean-field (or large ensemble) of conductances can be based on 
an aggregate observer that only involves the mean-field voltage of a population. Here, we also anticipate that the physical nature of 
the electrical circuit model will allow flexible designs of multi-scale and hierarchical adaptive controllers.

\end{itemize}

\section{Conclusion}

This paper has investigated the classical framework of model-reference adaptive control to design neuromodulatory controllers
in conductance-based neuronal models. A key message of the paper is that conductance-based models are linearly parameterised
in maximal conductance parameters, and that the adaptive control of individual conductances provides flexible adaptation principles
reminiscent of those observed in neurophysiology.  The proposed methodology makes use of the  adaptive observer recently proposed in \cite{burghi2021online}.
It fundamentally relies on the physical input-output properties of conductance-based models, namely the relative degree one property between currents
and voltages, and the contraction property of the internal (gating) dynamics. It is also decentralised, in the sense that the adaptive
controller of each node in a network only estimates local states and parameters based on local measurements, that is, the nodal voltage
and the voltage of presynaptic neurons. 

We have presented adaptive controllers to solve the two key control problems of reference tracking and disturbance rejection. 
We have also provided a simple illustration of the role of nodal adaptive control in a conceptual 
network inspired by the stomatogastric ganglion.

The results of this paper are preliminary in that they assume full knowledge of the individual conductance models and perfect
measurements of the voltage variables. The idealised results however suggest a strong potential for classical solutions
of adaptive control to provide neuromodulation principles in biophysical or artificial conductance-based networks.

\section*{Acknowledgements}

Raphael Schmetterling has received funding from an EPSRC Doctoral Training Programme. Thiago B. Burghi has received funding from the European Research Council under the ERC grant agreement FLEXNEURO n.716643.

The authors are thankful to Timothy O'Leary for fruitful discussions on these results. 

\appendix

\section{Model Parameters}
\label{sec:model_parameters} 

The simulation parameters are as follows. In Section \ref{sec:tracking}, the model parameters are given by
\[
\maxcond_r = 
\begin{pmatrix}
	120, & 0.1, & 2, & 0, & 80, & 0.4, & 2, & 0, & 0.1
\end{pmatrix}^\transp
\]
The input current $u_r = -2$ and $\kappa = 0.04$. The observer parameters are $\alpha = 0.0008$ and $\gamma = 2$.

In Section \ref{sec:rejection} the parameters are given by
\[
\maxcond = 
\begin{pmatrix}
	60, & 0.1, & 2, & 0, & 80, & 0.4, & 2, & 0, & 0.12
\end{pmatrix}^\transp
\]
and
\[
\maxcond_d = 
\begin{pmatrix}
	130, & 0.1, & 3.2, & 0, & 80, & 1, & 2, & 0, & 0.1
\end{pmatrix}^\transp
\]
The synapse has parameters $a_1 = 0.53$ and $a_2 = 0.18$. The activation function $\sigma_s$ has the form given in \eqref{eq:synaptic_time-constant}. The maximal conductance of the synapse is $\maxcond_{\syn} = 2.5$. The input currents are $\bar{u} = -2$ and $u_d = -1$, except in the period $t < 400$ when $u_d = -7.5$. This is to
delay the first burst of the disturbance neuron to better illustrate its impact. The observer parameters
are $\alpha = 0.001$ and $\gamma = 5$, and the controller parameter $\beta = 100$.

In Section \ref{sec:network} the parameters are given by
\[
\maxcond_1 = \maxcond_2 =
\begin{pmatrix}
	120, & 0.1, & 1.6, & 0, & 80, & 0.8, & 2, & 0, & 0.1
\end{pmatrix}^\transp
\]
and
\[
\maxcond_3 = 
\begin{pmatrix}
	60, & 0.1, & 2, & 0, & 30, & 0, & 1, & 0, & 0.1
\end{pmatrix}^\transp
\]
The maximal conductances of Neurons 4 and 5, $\maxcond_4$ and $\maxcond_5$, are the same as $\maxcond_d$ above. The synapses all have $a_1$, $a_2$ and $\sigma_s$ as in Section \ref{sec:rejection}, given above. The maximal conductances of the synapses are $\maxcond_{\syn,2,1} = \maxcond_{\syn,1,2} = 0.8$, $\maxcond_{\syn,5,4} = \maxcond_{\syn,4,5} = 0.6$ and $\maxcond_{\syn,1,3} = \maxcond_{\syn,5,3} = 8$. The gap junctions have conductance $\maxcond_\textrm{gap} = 0.004$.

The input currents are $\bar{u}_1 = \bar{u}_2 = -3.5$, $\bar{u}_3 = 38$ and $\bar{u}_4 = \bar{u}_5 = -3.2$, except in the period $t < 600$ when $\bar{u}_1 = -8$ and $\bar{u}_4 = -7$. This is
as the neurons in the HCO are at rest unless they are inhibited and then released from inhibition. This
release from inhibition generates a burst, which in turn inhibits and releases the other neuron in 
the HCO, and so on. The observer parameters are $\alpha = 0.0004$ and $\gamma = 5$, and the controller parameter $\beta = 100$.

All neurons simulated in this paper used the same values for the reversal potentials and the neuron capacitance. These values are listed in \cref{tab:reversal_potentials_and_c}.

\begin{table}
	\caption{Reversal potentials and neuron capacitance.}
	\normalsize
	\centering
	\begin{tabular}{|p{.8cm}|p{.8cm}|p{.8cm}|p{.8cm}|
			p{.8cm}|p{.8cm}|p{.8cm}|}
		\hline
		$\nernst_\Na$ & $\nernst_\Hcond$ &
		$\nernst_\Ca$ & $\nernst_\K$ &
		$\nernst_\syn$ & $\nernst_\Leak$ & $c$ 
		\\
		\hline
		$45$ & $-43$ & $120$ & $-90$ & $-90 $ & $-55$ & $0.1$
		\\
		\hline
	\end{tabular}
	\label{tab:reversal_potentials_and_c} 
\end{table}

\section{Gating Dynamics}
\label{sec:gating_dynamics} 

All neurons share the same gating variables \eqref{eq:current_cb}.
The activation and time-constant functions for each gate are provided below. First note that some of these functions have been defined in terms of functions $\alpha_\ion(v)$ and $\beta_\ion(v)$:
\begin{align*}
	\alpha_{m_{\Na}} &= \frac{-0.025(v+40)}{\exp(-(v+40)/10)-1}
	& \beta_{m_{\Na}} &= \exp(-(v+65)/18) \\
	\alpha_{h_{\Na}} &= 0.0175\exp(-(v+65)/20)
	& \beta_{h_{\Na}} &= \frac{0.25}{1 + \exp(-(v+35)/10)} \\
	\alpha_{m_{\K}} &= \frac{0.0025(v+55)}{1 - \exp(-(v+55)/10)}
	& \beta_{m_{\K}} &= 0.03125\exp(-(v+65)/80) \\
	\alpha_{m_{\Hcond}} &= \exp(-14.59-0.086v)
	& \beta_{m_{\Hcond}} &= \exp(-1.87+0.0701v)
\end{align*}
The activation functions are as follows:
\begin{align*}
	\sigma_{m, \Na} &= \frac{\alpha_{m_{\Na}}(v)}{\alpha_{m_{\Na}}(v)+\beta_{m_{\Na}}(v)}
	& \sigma_{h, \Na} &= \frac{\alpha_{h_{\Na}}(v)}{\alpha_{h_{\Na}}(v)+\beta_{h_{\Na}}(v)} \\
	\sigma_{m, \K} &= \frac{\alpha_{m_{\K}}(v-10)}{\alpha_{m_{\K}}(v-10)+\beta_{m_{\K}}(v-10)}
	& \sigma_{m, \Hcond} &= \frac{\alpha_{m_{\Hcond}}(v)}{\alpha_{m_{\Hcond}}(v)+\beta_{m_{\Hcond}}(v)} \\
	\sigma_{m, \A} &= \frac{1}{1+\exp(-(v+90)/8.5)}
	& \sigma_{h, \A} &= \frac{1}{1+\exp((v+78)/6)} \\
	\sigma_{m, \T} &= \frac{1}{1+\exp(-(v+57)/6.2)}
	& \sigma_{h, \T} &= \frac{1}{1+\exp((v+81)/4.03)} \\
	\sigma_{m, \Lcond} &= \frac{1}{1+\exp(-(v+55)/3)}
	& \sigma_{m,\KIR} &= \frac{1}{1+\exp((v+107.9)/9.7)}
\end{align*}
The time-constant functions are as follows:
\begin{align*}
	\tau_{m, \Na} &= \frac{1}{0.2(\alpha_{m_{\Na}}(v)+\beta_{m_{\Na}}(v))}
	& \tau_{h, \Na} &= \frac{1}{0.2(\alpha_{h_{\Na}}(v)+\beta_{h_{\Na}}(v))} \\
	\tau_{m, \K} &= \frac{1}{0.2(\alpha_{m_{\K}}(v-10)+\beta_{m_{\K}}(v-10))}
	& \tau_{m, \Hcond} &= \frac{1}{\alpha_{m_{\Hcond}}(v)+\beta_{m_{\Hcond}}(v)} \\
	\tau_{m, \Lcond} &= 72\exp(-(v+45)^2/400)+6
\end{align*}
\begin{align*}
	\tau_{m, \T} &= 0.612 + \frac{1}{\exp(-(v+131.6)/16.7)+\exp((v+16.8)/18.2)} \\
	\tau_{m, \A} &= 0.37 + \frac{1}{0.2(\exp((v+35.82)/19.697)+\exp((v+79.69)/-12.7))}
\end{align*}

We also have
\[
\tau_{h, \A} = 
\begin{cases}
	\frac{1}{0.2(\exp((v+46.05)/5)+\exp((v+238.4)/-37.45))}& \text{if } v < -63,\\
	19              & \text{otherwise}
\end{cases}
\]
and
\[
\tau_{h, \T} = 
\begin{cases}
	\exp((v+467)/66.6)& \text{if } v < -80,\\
	\exp(-(v+21.88)/10.2)+28              & \text{otherwise}.
\end{cases}
\]






\bibliographystyle{ieeetran}
\bibliography{ifac-references,ifac-refs-2}

\end{document}